\documentclass[11pt,a4paper]{article}

\usepackage[utf8]{inputenc}
\usepackage[T1]{fontenc}
\usepackage[protrusion=true,expansion=false]{microtype}
\usepackage{geometry}
\usepackage{amsmath,amssymb,amsthm}

\usepackage{booktabs}
\usepackage{tabularx}
\usepackage{array}
\usepackage{multirow}

\usepackage{listings}
\usepackage{xcolor}

\definecolor{codebg}{RGB}{247,247,247}
\definecolor{codeframe}{RGB}{180,180,180}
\definecolor{keyword}{RGB}{0,0,180}
\definecolor{comment}{RGB}{100,100,100}
\definecolor{string}{RGB}{150,0,0}
\definecolor{deepblue}{RGB}{31,56,100}
\definecolor{midblue}{RGB}{46,92,168}

\lstdefinestyle{codestyle}{
  backgroundcolor=\color{codebg},
  frame=single,
  framerule=0.5pt,
  rulecolor=\color{codeframe},
  basicstyle=\ttfamily\footnotesize,
  keywordstyle=\color{keyword}\bfseries,
  commentstyle=\color{comment}\itshape,
  stringstyle=\color{string},
  breaklines=true,
  breakatwhitespace=false,
  tabsize=2,
  captionpos=b,
  numbers=left,
  numberstyle=\tiny\color{comment},
  numbersep=6pt,
  showstringspaces=false,
  xleftmargin=1.5em,
  xrightmargin=0.5em,
}

\lstdefinelanguage{Rego}{
  morekeywords={package,import,default,allow,deny,true,false,if,else,contains,every,some},
  sensitive=true,
  morecomment=[l]{\#},
  morestring=[b]",
}

\lstdefinelanguage{DOT}{
  morekeywords={digraph,graph,node,edge,subgraph},
  sensitive=true,
  morecomment=[l]{//},
  morestring=[b]",
}

\usepackage{graphicx}
\usepackage{float}
\usepackage{caption}
\usepackage[numbers,sort&compress]{natbib}
\usepackage[colorlinks=true,
            linkcolor=deepblue,
            citecolor=midblue,
            urlcolor=midblue]{hyperref}
\usepackage{cleveref}

\newtheorem{definition}{Definition}

\newcommand{\MCP}{\textsc{mcp}}
\newcommand{\OPA}{\textsc{opa}}
\newcommand{\POMDP}{\textsc{pomdp}}
\newcommand{\EPA}{\textsc{epa}}
\newcommand{\BOLA}{\textsc{bola}}
\newcommand{\RBAC}{\textsc{rbac}}
\newcommand{\LLM}{\textsc{llm}}
\newcommand{\HitL}{\textsc{h-i-t-l}}
\newcommand{\PoC}{\textsc{poc}}

\usepackage{appendix}

\usepackage{titlesec}
\titleformat{\section}{\large\bfseries\color{deepblue}}{\thesection}{1em}{}
\titleformat{\subsection}{\normalsize\bfseries\color{midblue}}{\thesubsection}{1em}{}
\titleformat{\subsubsection}{\normalsize\bfseries}{\thesubsubsection}{1em}{}

\usepackage{mdframed}
\mdfdefinestyle{abstractbox}{
  linecolor=midblue,
  linewidth=1pt,
  leftmargin=0.5cm,
  rightmargin=0.5cm,
  innerleftmargin=0.5cm,
  innerrightmargin=0.5cm,
  innertopmargin=0.3cm,
  innerbottommargin=0.3cm,
  backgroundcolor=blue!3,
}

\title{%
  \textbf{\large From CRUD to Autonomous Agents:}\\[0.4em]
  \textbf{Formal Validation and Zero-Trust Security for\\
  Semantic Gateways in AI-Native Enterprise Systems}
}

\author{%
  Ignacio Peyrano\\
  \normalsize Universidad Austral\\
  \normalsize Rosario, Argentina\\
  \normalsize \texttt{peyranoignacio@gmail.com}
}

\date{April 2026}

\begin{document}

\maketitle
\thispagestyle{empty}

\begin{mdframed}[style=abstractbox]
\textbf{Abstract.}
Enterprise software engineering is undergoing a foundational structural transition. The
industry is rapidly shifting away from deterministic architectures based on Create, Read,
Update, Delete (CRUD) operations, Representational State Transfer (REST) contracts, and
predefined execution flows. In their place, organizations are adopting AI-native systems
where large language models function as cognitive orchestrators capable of interpreting
ambiguous intentions, dynamically composing computational tools, and executing complex,
multi-step plans. However, this transition introduces a profound architectural and security
tension. While the probabilistic nature of large language models enables a superior, highly
flexible semantic interaction layer, it simultaneously weakens classical mechanisms for
software validation, granular access control, and formal testing.

This research proposes the design, formal mathematical validation, and empirical evaluation
of a \textit{Semantic Gateway} architecture governed by the Model Context Protocol (\MCP{}).
Positioned as a robust, stateful alternative to exposing granular REST topologies directly
to artificial agents, the Semantic Gateway reframes the enterprise \textsc{api} as a
semantic surface where tools are dynamically discovered, authorized, and executed based on
intent, context, and rigorous policy enforcement. The central scientific contribution is
built upon a fundamental paradigm shift: \textit{autonomous agents must not be validated as
traditional software modules nor as simple API consumers, but as stochastic state-transition
systems whose behavior must be abstracted, fuzzed, and audited through enabled-tool graphs.}

To operationalize this principle, the proposed architecture introduces a comprehensive
three-layer Zero-Trust security model comprising a pre-inference Semantic Firewall,
deterministic Tool-Level Role-Based Access Control (\RBAC{}), and out-of-band Cryptographic
Human-in-the-Loop (\HitL{}) approval. Furthermore, this research adapts
Enabledness-Preserving Abstractions (\EPA{}s) and greybox semantic fuzzing
methodologies---historically utilized for the formal verification of immutable blockchain
smart contracts---to dynamically audit agent behavior in enterprise environments. Extensive
experimental results demonstrate an \textbf{84.2\% reduction} in incidental code. Through
the execution of 500,000 multi-turn fuzzing sequences, the methodology achieved a
\textbf{100\% discovery rate} of hidden, unauthorized state transitions, proving that
dynamic formal verification is not only viable but strictly necessary for the secure
deployment of agentic enterprise systems.
\end{mdframed}

\vspace{0.5em}

\tableofcontents
\newpage

\section{Introduction}
\label{sec:intro}

The evolution of enterprise software architecture has historically been driven by the
imperative to manage system complexity through deterministic, highly predictable
abstractions. From structured monolithic systems to Service-Oriented Architectures and
microservices proliferation, the underlying engineering paradigm has remained remarkably
stable: software design predicated on static rules, rigid interface contracts, and entirely
predictable execution flows. However, the integration of autonomous
agentic artificial intelligence systems into enterprise environments precipitates the rapid
collapse of this structural rigidity. The transition from systems of explicit instruction
to systems of intent-based execution demands a fundamental rethinking of how digital
systems interact, authenticate, and maintain verifiable security boundaries.

For decades, dominant communication architectures---namely REST, GraphQL, and Remote
Procedure Call---have operated on a stable assumption: the consumer possesses \textit{a
priori} knowledge of the backend topology. When the consumer is no longer a human
developer or a deterministic software service, but an autonomous agent utilizing a large
language model (\LLM{}) capable of probabilistic reasoning, this assumption fails
completely.

Exposing traditional API endpoints directly to \LLM{}s induces severe architectural
friction. Foremost is \textit{context window bloat}: exposing an agent to hundreds of
granular endpoints via extensive OpenAPI specifications rapidly consumes the model's token
limits~\cite{contexttokens}. Furthermore, the inherently stochastic nature of the \LLM{}
introduces novel threat vectors that traditional API gateways were not designed to
mitigate, including multi-turn prompt injections, context poisoning, and Broken Object
Level Authorization (\BOLA{}) exploitation achieved through hallucinated combinations of
otherwise benign tools~\cite{owasp_top10}.

To resolve these escalating architectural tensions, this research introduces the
comprehensive design and formal validation of a \textit{Semantic Gateway}. Rather than
acting as a simple, stateless routing layer, the Semantic Gateway operates as an epistemic
and operational frontier between the stochastic agent and the deterministic enterprise
backend. Crucially, this research pioneers the importation of formal
verification techniques from the domain of blockchain smart contracts. By mathematically
modeling the agent's environment as a Partially Observable Markov Decision Process
(\POMDP{}) and applying Enabledness-Preserving Abstractions alongside greybox semantic
fuzzing, the system transforms the unpredictable behavior of large language models into a
mathematically auditable, finite graph of state transitions.

\section{Background and Related Work}
\label{sec:background}

To establish the scientific gap addressed by the Semantic Gateway, it is necessary to
critically examine the current landscape of API contracts, agent orchestration frameworks,
artificial intelligence security policies, and formal verification methodologies. The
existing literature demonstrates rapid advancements in isolated domains, yet reveals a
profound lack of cohesive, formal methodologies for securing emergent behaviors of
autonomous agents.

\subsection{REST, GraphQL, and the Limits of Deterministic Contracts}

The Representational State Transfer architecture established a resource-oriented paradigm
that remains foundational for web interoperability, while GraphQL introduced a strongly
typed alternative allowing clients to declare the exact shape of required data. Both
models succeed brilliantly under the premise of deterministic consumption. However, in
agentic scenarios, they generate immense semantic friction: the \LLM{} is forced to act as
an impromptu network compiler, translating high-level natural language objectives into
highly specific HTTP methods and nested JSON payloads~\cite{contexttokens}. This cognitive
load significantly increases the probability of tool hallucination and execution failure,
proving that REST is an insufficient abstraction layer for cognitive agents.

\subsection{The Model Context Protocol and Tool Calling}

The Model Context Protocol (\MCP{}), released by Anthropic in late 2024, has rapidly
emerged as the open standard to unify how AI models connect to external data sources and
computational tools~\cite{mcp_spec}. Operating over a standardized JSON-RPC
implementation, \MCP{} defines the schema for exposing functional tools to a model,
allowing the agent to dynamically discover capabilities at runtime. However, researchers
have identified critical security limitations: the protocol itself intentionally delegates
authorization to the underlying transport layer, meaning seamless interoperability does not
inherently guarantee governance or cryptographic security~\cite{mcp_security}.

\subsection{Agent Orchestration: LangChain versus LangGraph}

The orchestration of \LLM{} agents has evolved significantly, shifting from linear,
stateless pipelines exemplified by LangChain, toward graph-based, stateful architectures
like LangGraph~\cite{langchain_vs_langgraph}. LangGraph natively models multi-agent
workflows as cyclical state machines, where nodes represent discrete units of work and
edges represent conditional routing logic based on the model's output~\cite{langgraph_docs}.
While LangGraph effectively manages the operational flow and memory persistence of complex
agentic loops, it does not inherently provide any formal verification of the resulting
state space~\cite{langgraph_decision}.

\subsection{AI Security Frameworks: OWASP GenAI and NIST AI RMF}

The security posture of \LLM{} applications is currently guided by prominent frameworks
such as the OWASP Top 10 for Large Language Model Applications~\cite{owasp_top10} and the
NIST Artificial Intelligence Risk Management Framework~\cite{nist_ai_rmf}. OWASP highlights
critical vulnerabilities including Prompt Injection, Insecure Output Handling, and
Excessive Agency, explicitly warning against granting models unfettered access to
enterprise APIs. However, neither framework offers a concrete, mathematical methodology
for dynamically validating the execution boundaries of an agent at runtime~\cite{nist_agentic}.

\subsection{Open Policy Agent and Policy as Code}

The Open Policy Agent (\OPA{}) is a general-purpose, open-source policy engine that
utilizes a highly optimized declarative language known as Rego to decouple policy
decision-making from the underlying application's business logic~\cite{opa_docs}. By
utilizing this engine, organizations can define strict \RBAC{} and Attribute-Based Access
Control (ABAC) rules as immutable code~\cite{opa_use_cases}. In the context of AI agents,
\OPA{} can intercept a model's request to execute a tool and deterministically evaluate
whether the agent possesses the requisite cryptographic permissions, ensuring that
authorization remains a deterministic function outside the model's probabilistic
influence~\cite{opa_ai_guardrail}.

\subsection{Formal Verification, Smart Contract Fuzzing, and EPAs}

Due to the strict immutability of blockchain environments, decentralized smart contracts
are routinely subjected to rigorous formal verification. Tools such as Echidna utilize
property-based testing and greybox fuzzing to generate hundreds of thousands of random
transaction sequences, aiming to violate predefined system invariants~\cite{echidna_issta}.
A highly effective abstraction model is the Enabledness-Preserving Abstraction (\EPA{}),
which quotients an infinite programmatic state space into finite equivalence classes
grouped based on the specific set of methods they currently enable~\cite{epa_contracts}.
This paper pioneers the direct adaptation of \EPA{}s to AI agents, treating the agent's
available tools as the set of abstract methods, and utilizing targeted semantic fuzzing to
exhaustively map the agent's emergent behavioral graph.

\section{Problem Statement}
\label{sec:problem}

As global enterprises aggressively integrate agentic AI into their core operations, they
confront an asymmetrical and highly unpredictable risk landscape. Traditional software
components are inherently deterministic; given the exact same inputs and system state,
they produce the exact same outputs. Autonomous agents driven by \LLM{}s, however, are
fundamentally stochastic systems.

\begin{mdframed}[style=abstractbox]
\textbf{Central Thesis.}
\textit{Autonomous agents must not be validated as traditional software nor as simple API
consumers, but as stochastic state-transition systems whose behavior must be abstracted,
fuzzed, and audited through enabled-tool graphs.}
\end{mdframed}

\vspace{0.5em}

The problem manifests along four critical dimensions:

\textbf{Contextual Vulnerability and Memory Tainting.} An autonomous agent routinely
ingests data from external, untrusted sources. During this process, the agent may
inadvertently read a malicious instruction embedded within the text---an attack vector
known as \textit{indirect prompt injection}. If the agent's working memory becomes tainted,
it may subsequently invoke a write-access tool using the compromised context. Because
traditional perimeter defenses authenticate the identity of the agent rather than the
origin of the prompt context, the malicious mutation is executed, bypassing standard
security measures entirely.

\textbf{Semantic Broken Object Level Authorization (\BOLA{}).} This vulnerability occurs
when an application correctly authenticates a user but completely fails to verify that the
user possesses authorization to access or modify a specific object identifier requested in
the API call~\cite{bola_pynt}. An agent, capable of generating synthetic parameters or
hallucinating object identifiers during its reasoning process, can inadvertently or
maliciously attempt to access unauthorized records~\cite{bola_f5}.

\textbf{Emergent, Unpredictable Transitions.} A severe vulnerability may only manifest
when a state-modifying tool is immediately followed by a privilege-escalating tool---an
emergent interaction sequence that static code analysis and standard integration tests
cannot reliably predict or cover.

\textbf{The API Paradox.} Exposing highly granular REST endpoints to a language model
leads to massive context window bloat. Conversely, creating monolithic macro-tools designed
to reduce token consumption leads to tightly coupled, inflexible systems that defeat the
entire purpose of dynamic, modular agent composition~\cite{contexttokens}.

\section{Semantic Gateway Architecture}
\label{sec:architecture}

To bridge the architectural chasm between stochastic agent reasoning and deterministic
enterprise execution, this research proposes the comprehensive design of the Semantic
Gateway. Governed by \MCP{}, the Semantic Gateway acts as a stateful, intelligent
infrastructure layer that completely abstracts heterogeneous enterprise backend systems
into a unified, highly governed semantic interface.

\subsection{Architectural Components}

The architecture is composed of interconnected layers designed to process, sanitize, route,
and audit agentic interactions.

\textbf{Intent Intake Layer.} Serves as the primary ingress point, ingesting unstructured
natural language requests directly from users or semi-structured JSON payloads generated
by upstream autonomous systems. This layer decouples the user's ultimate objective from
the specific technical mechanisms required to achieve it.

\textbf{Semantic Normalizer and Vectorizer.} Converts the raw linguistic request into a
highly dense mathematical vector embedding, normalizing ambiguous human terminology into a
canonical organizational vocabulary.

\textbf{Semantic Firewall.} Operating entirely pre-inference, this layer utilizes highly
efficient micro-models, such as DeBERTa-v3, to analyze the embedded intent for adversarial
patterns, prompt injection signatures, or blatant policy violations before the costly and
vulnerable primary \LLM{} is engaged.

\textbf{Embedding Router and Tool Registry.} Calculates the cosine similarity between the
embedded intent and the semantic descriptions of all available \MCP{} tools. It
intelligently filters the global tool registry to construct an optimal, highly relevant
subset of tools that comfortably fits within the agent's context window.

\textbf{Chain-of-Thought Planner.} The core cognitive engine: the \LLM{} synthesizes a
multi-step execution plan, deciding which tools to call and in what exact sequence.
However, the agent's plan is treated merely as a highly suspect suggestion.

\textbf{Policy Enforcement Point.} Prior to execution, the proposed action, the target
resource, and the generated parameters are evaluated deterministically by an \OPA{}
instance. If the action violates the strict \RBAC{} logic, it is instantly blocked.

\textbf{Cryptographic Human-in-the-Loop Gate.} If the \OPA{} flags the requested tool as
highly critical or irreversible, execution is suspended. A secure cryptographic challenge
detailing the intent and parameters is generated and pushed to a human operator for
out-of-band signature approval, ensuring that autonomous agents cannot unilaterally execute
destructive actions.

\textbf{Audit Ledger.} Maintains an immutable, hash-chained record of the original intent,
the generated plan, the evaluated policy, the applied tool parameters, and the final state
mutation, providing perfect forensic traceability for the autonomous system.

\begin{figure}[H]
  \centering
  \includegraphics[width=0.62\textwidth]{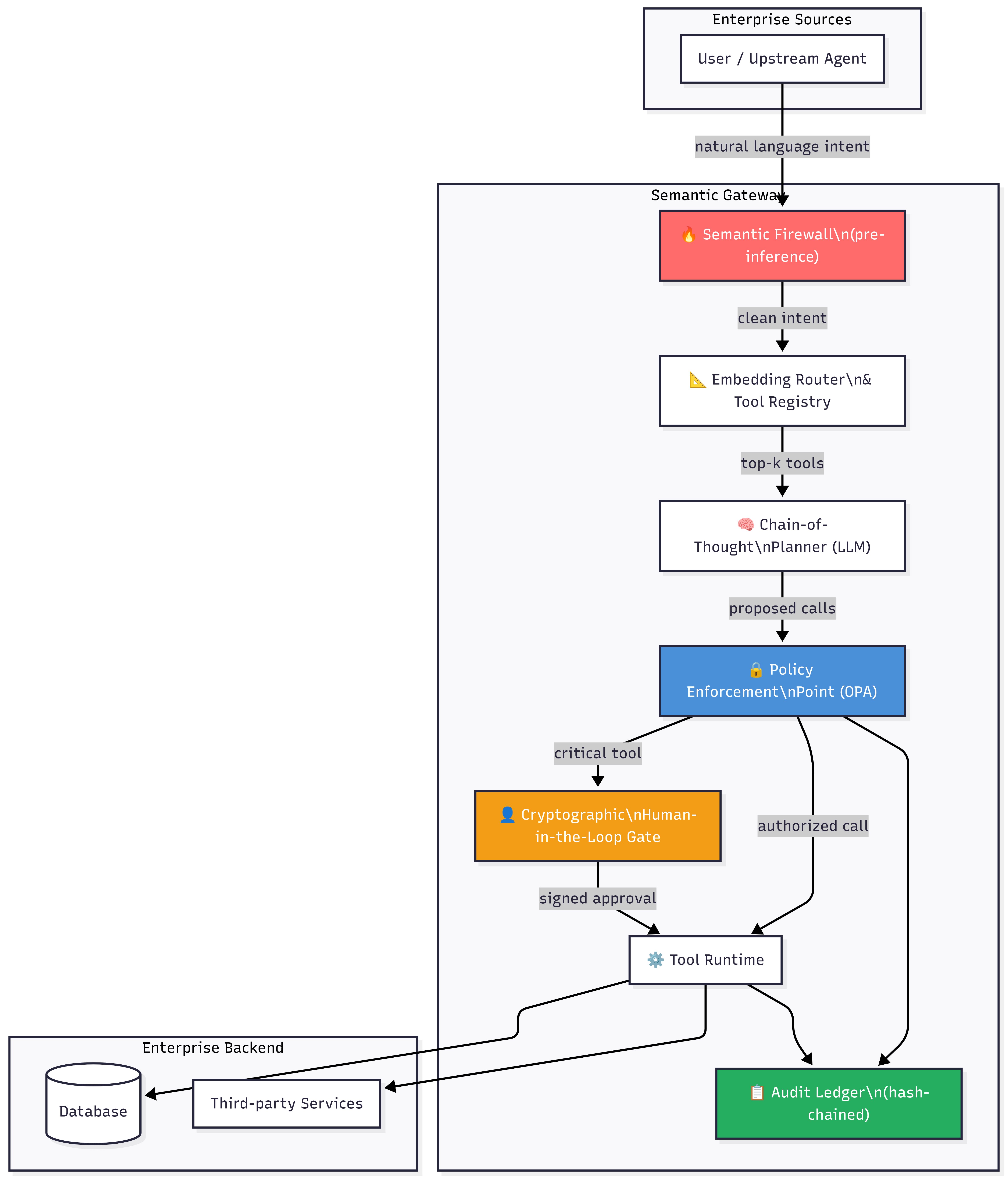}
  \caption{Semantic Gateway architecture. The intent flows from enterprise sources through the Semantic Firewall, Embedding Router, Chain-of-Thought Planner, and Policy Enforcement Point before reaching the Tool Runtime and Audit Ledger.}
  \label{fig:architecture}
\end{figure}

\Cref{fig:architecture} illustrates the complete component flow of the Semantic Gateway.

\section{Zero-Trust Security Model}
\label{sec:zerotrust}

The foundational security philosophy of the Semantic Gateway relies on a singular,
uncompromising premise: \textit{the large language model is inherently untrusted}. The
model possesses operational agency to plan and suggest, but absolutely zero authoritative
permission to execute. To enforce this strict separation, the architecture implements a
defense-in-depth model comprising three distinct, overlapping layers of verification.

\subsection{Layer 1: Semantic Firewall and Cognitive Isolation}

Positioned significantly before the main \LLM{} inference phase, the Semantic Firewall
serves to filter adversarial input and meticulously manage context safety. A primary
function of this layer is \textit{Context Isolation}: the firewall enforces strict
attention barriers within the transformer's processing layers, ensuring that user-provided
text or external data cannot overwrite the system's foundational meta-prompt or core
security directives. Furthermore, the firewall implements \textit{Taint-Aware Memory} using
advanced information flow tracking. Any data retrieved from an untrusted source is
immediately tagged in the agent's memory state as tainted. If the agent subsequently
attempts to pass this tainted context into a critical mutation tool, the Semantic Firewall
deterministically aborts the invocation.

\subsection{Layer 2: Tool-Level RBAC and Micro-segmentation}

Traditional REST architectures rely heavily on JSON Web Tokens validating broad access to
entire network routing paths. The Semantic Gateway abolishes these routes entirely,
shifting the security boundary to the tool level. It maps Non-Human Identities directly
to specific tool schemas defined by \MCP{}. When the language model determines it needs to
call a specific tool, the request is instantly routed to the \OPA{}. This engine evaluates
a strict Rego policy that cross-references the agent's assigned non-human role against the
granular preconditions required by the tool. This micro-segmentation enforces the principle
of least privilege at the function level.

\subsection{Layer 3: Cryptographic Human-in-the-Loop Integration}

For high-stakes, irreversible operations, allowing full autonomous execution introduces an
unacceptable level of operational risk. The Semantic Gateway integrates a stringent
protocol inspired by the IETF drafts for Confirmation with Human in the Loop
(\textsc{cheq})~\cite{cheq_draft} and the Agent Credential Attestation Protocol
(\textsc{acap})~\cite{acap_draft}. When a critical tool is invoked, the gateway initiates
a speculative suspension, pausing the execution thread entirely. It then generates an
immutable evidence package by hashing the proposed action, the exact parameters, and the
agent's recorded chain-of-thought. This challenge is pushed out-of-band to an authorized
human's secure device. The gateway resumes execution only if the signature verification
proves mathematically sound.

\begin{figure}[H]
  \centering
  \includegraphics[width=0.96\textwidth]{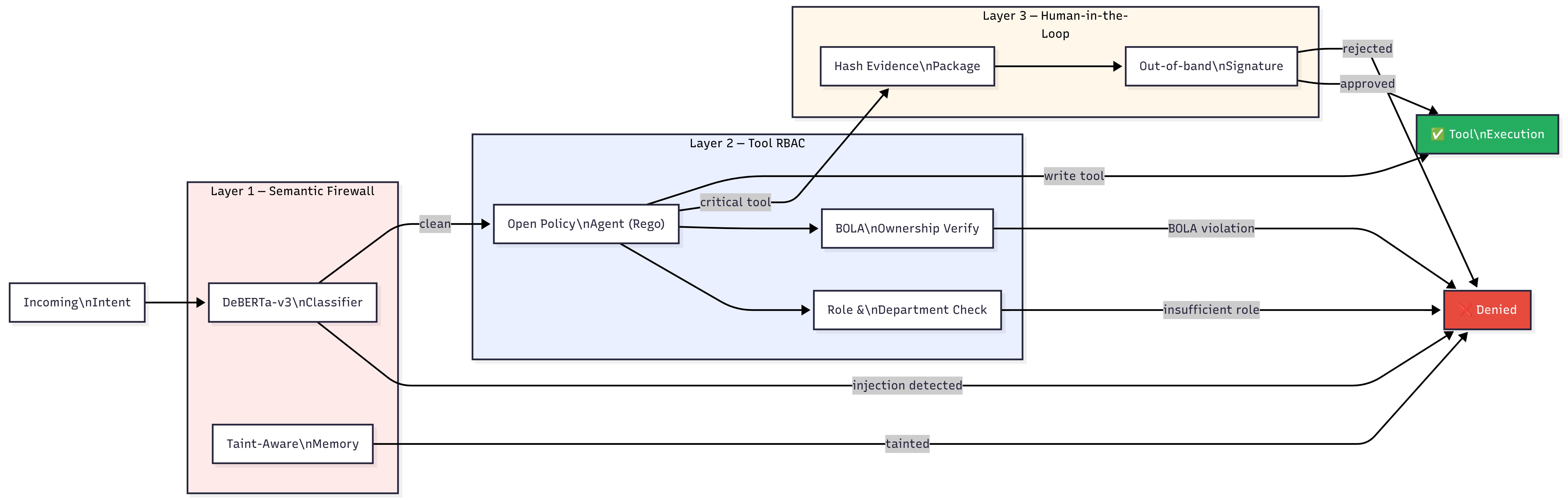}
  \caption{Three-layer Zero-Trust security model. Each incoming intent must survive Layer~1 (Semantic Firewall), Layer~2 (OPA-based Tool RBAC with BOLA ownership verification), and Layer~3 (Cryptographic Human-in-the-Loop) before any tool execution is permitted.}
  \label{fig:zerotrust}
\end{figure}

\Cref{fig:zerotrust} depicts the three-layer defense-in-depth model and the decision paths for denial and execution.

\section{Formal Model}
\label{sec:formal}

The rigorous validation and mathematical auditing of the Semantic Gateway requires
expressing its core operations, routing decisions, and state transitions through formal
mathematical constructs.

\subsection{Vector Space Semantic Routing}

The semantic routing mechanism is formally defined within a continuous, highly dimensional
vector space $\mathcal{V} \subset \mathbb{R}^n$. Given a natural language user intent $q$
and a pre-trained embedding model $\varphi$, the intent vector is calculated as:
\begin{equation}
  \mathbf{v}_q = \varphi(q)
\end{equation}

Let the set $\{\mathbf{v}_{t_i}\}_{i=1}^{|T|}$ represent the mathematical semantic
embeddings of the descriptive schemas of all available \MCP{} tools. The relevance of any
specific tool $t_i$ to the current intent is computed utilizing the Cosine Similarity
metric:
\begin{equation}
  \mathrm{sim}(\mathbf{v}_q, \mathbf{v}_{t_i})
    = \frac{\mathbf{v}_q \cdot \mathbf{v}_{t_i}}
           {\|\mathbf{v}_q\| \cdot \|\mathbf{v}_{t_i}\|}
\end{equation}

To actively prevent the degradation of reasoning capabilities caused by context window
bloating, the gateway resolves a subset optimization challenge, mathematically analogous to
the classical Knapsack problem. It selects the optimal set of tools $T^*$ subject to a
strict maximum token limit $L_{\max}$:
\begin{equation}
  T^* = \underset{\substack{T \subseteq T_{\text{all}} \\
                             \sum_{t_i \in T} \mathrm{tokens}(t_i) \leq L_{\max}}}
        {\operatorname{argmax}} \sum_{t_i \in T} \mathrm{sim}(\mathbf{v}_q, \mathbf{v}_{t_i})
\end{equation}

\subsection{Planning as a Partially Observable Markov Decision Process}

The agent's complex, multi-step tool orchestration process is formally modeled as a
Partially Observable Markov Decision Process (\POMDP{}). In this construct, the actual
business state $s \in \mathcal{S}$ is only partially observable to the agent, which yields
periodic observations $o \in \mathcal{O}$ via the execution outputs of tools. The action
space $\mathcal{A}$ comprises the set of tools currently permitted by the policy engine.
The probability of the agent achieving a specific goal $g$, given a starting intent $q$
and an initial context $h_0$, relies heavily on the \LLM{}'s internal policy $\pi_\theta$,
parameterized by its neural network weights $\theta$:
\begin{equation}
  P(g \mid q, h_0)
    = \sum_{\tau} \pi_\theta(a_t \mid h_t)
      \cdot P(o_{t+1} \mid s_t, a_t)
      \cdot \delta(s_{t+1}, g)
\end{equation}
where $h_t$ represents the historical, dynamic context comprising all previous actions and
observations. Because the policy $\pi_\theta$ is inherently stochastic and highly
vulnerable to adversarial drift through prompt injection, the mathematical integration of
deterministic \OPA{} rules is critical. This policy matrix ensures that the transition
probability $P(a_i \mid \OPA{}_{\text{deny}})$ is strictly forced to $0$, permanently
overriding the model's probabilistic likelihood of failure.

\subsection{Enabledness-Preserving Abstractions}

To formally verify the security of the entire agentic system, the concrete, potentially
infinite state of the enterprise application is abstracted into a highly structured graph
using \EPA{}s. This mechanism groups concrete programmatic states into abstract equivalence
classes based entirely on which specific tools are currently enabled by the strict
preconditions of the \RBAC{} matrix~\cite{epa_contracts}.

\begin{definition}[\EPA{} of an Agentic System]
Formally, the agent's entire behavioral abstraction is defined as a tuple
$\mathcal{A} = (T, \tilde{S}, \tilde{s}_0, \tilde{\delta})$, where:
\begin{itemize}
  \item $T$ is the complete set of all available tools within the ecosystem.
  \item $\tilde{S} \subseteq 2^T$ is the power set of $T$, where each distinct abstract
        state $\tilde{s}$ represents a unique, valid combination of currently permitted
        tools.
  \item $\tilde{s}_0 \in \tilde{S}$ is the initial, default permission state of the agent.
  \item $\tilde{\delta}: T \times \tilde{S} \to \tilde{S}$ is the critical transition
        function detailing how invoking a tool $t$ while residing in abstract state
        $\tilde{s}$ modifies the underlying context to reach a new abstract state.
\end{itemize}
\end{definition}

A valid state transition is expressed as:
\begin{equation}
  \tilde{\delta}(t, \tilde{s}) = \tilde{s}'
  \iff
  \exists\, c \in \mathcal{C} \;:\;
    \bigl[c \models \varphi(\tilde{s})
    \;\wedge\;
    \mathrm{Exec}(t, \mathit{params}) \models c'\bigr]
  \;\wedge\;
  \tilde{s}' = \mathrm{enabled}(c')
\end{equation}

\begin{figure}[H]
  \centering
  \includegraphics[width=0.68\textwidth]{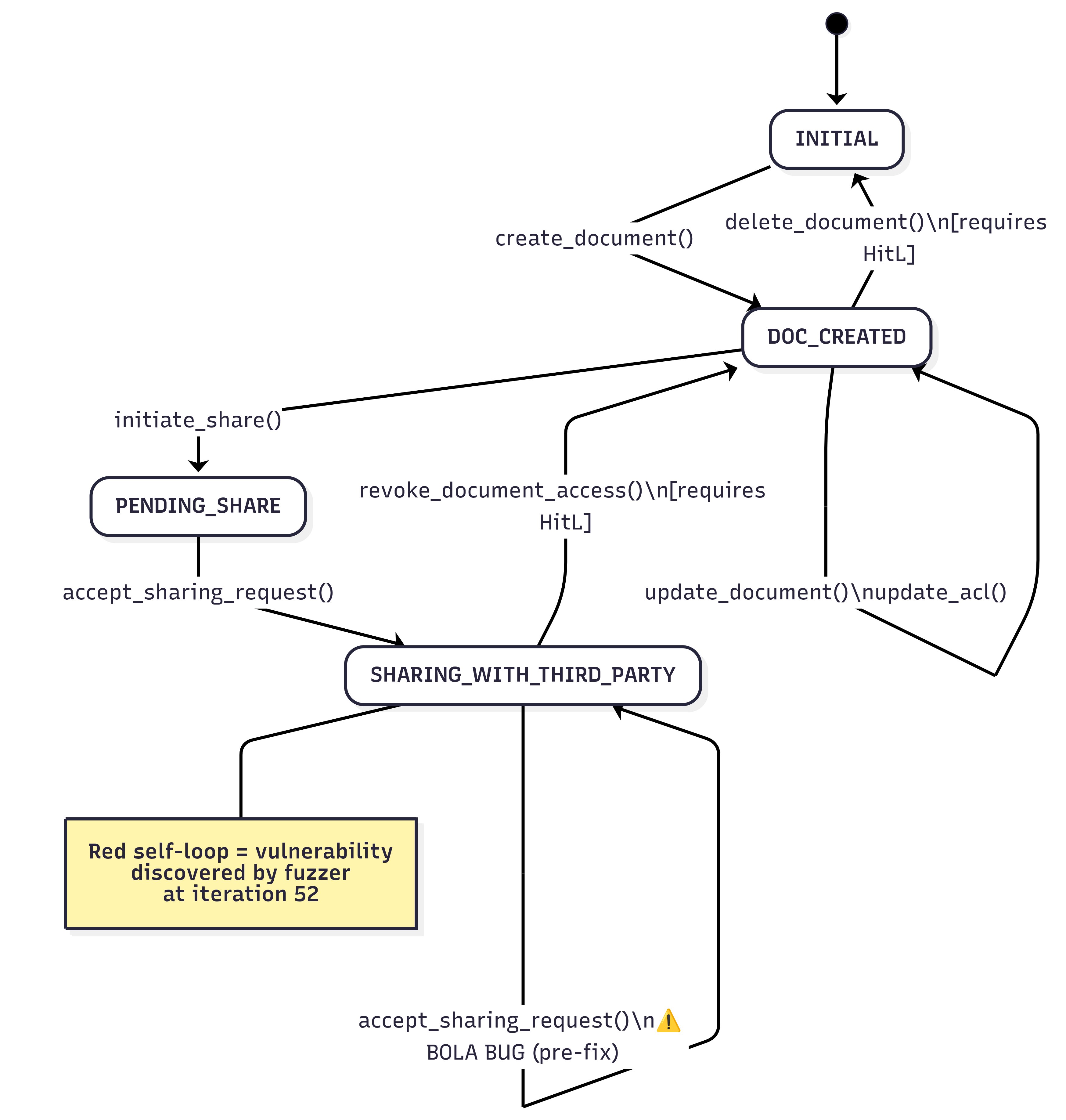}
  \caption{Enabledness-Preserving Abstraction (\EPA{}) graph for the Document Sharing workflow. The yellow self-loop on \texttt{SHARING\_WITH\_THIRD\_PARTY} represents the \BOLA{} vulnerability discovered by the semantic fuzzer at iteration~52. The transition is absent in the fixed \EPA{} after the Rego policy amendment.}
  \label{fig:epa}
\end{figure}

This formula dictates that a valid state transition exists if and only if, under the
current context $c$ that perfectly satisfies all system invariants and the specific
predicates of state $\tilde{s}$, the successful execution of tool $t$ results in a new
context that flawlessly satisfies the predicates defining abstract state $\tilde{s}'$.
This mathematical foundation is what allows the system to be rigorously fuzzed. \Cref{fig:epa} shows the concrete \EPA{} graph for the Document Sharing workflow evaluated in this paper, including the discovered vulnerability.

\section{Semantic Fuzzing Methodology}
\label{sec:fuzzing}

Traditional unit testing is entirely inadequate for securing stochastic autonomous agents,
as it cannot account for emergent reasoning patterns or unpredictable tool combinations.
Therefore, this research adapts advanced greybox fuzzing methodologies---proven highly
effective in discovering devastating vulnerabilities within immutable Ethereum smart
contracts via analysis engines like Echidna~\cite{echidna_github,echidna_issta} and
Halmos---into a bespoke semantic fuzzing framework designed specifically for the Semantic
Gateway.

\subsection{The Semantic Fuzzer Harness}

The automated fuzzer harness is designed to autonomously inject hundreds of thousands of
stochastic intents, highly mutated structural parameters, and complex adversarial context
sequences directly into the gateway architecture. It operates by systematically attempting
to falsify predefined mathematical assertions regarding the system's security posture. For
example, the harness dynamically constructs and executes test cases structured as follows:

\begin{lstlisting}[language=C, caption={Semantic invariant test analogous to Echidna harness}, label={lst:harness}]
function from_state_A_to_B_by_tool_K(params) {
    require(precondition_state_A);
    execute_tool_K(params);
    assert(!preconditions_state_B);
}
\end{lstlisting}

If the fuzzer algorithm discovers a specific combination of adversarial parameters or
prompt sequences that successfully falsifies the terminal assertion, it mathematically
proves the existence of a highly dangerous, undocumented state transition from abstract
state $\tilde{s}_A$ to abstract state $\tilde{s}_B$ via the execution of tool $t_K$.

\subsection{State Space Optimization Heuristics}

Because the potential abstract state space of the agentic ecosystem scales exponentially,
represented as $O(2^{|T|})$ where $|T|$ is the number of available tools, exhaustive,
brute-force fuzzing risks catastrophic combinatorial explosion. To ensure the fuzzing
process remains computationally tractable, three critical reduction heuristics are applied:

\textbf{Reduce\_True.} Tools categorized as purely read-only inherently lack restrictive
state-modifying preconditions and are universally available across all states. These tools
are globally abstracted out of the transition matrix, effectively halving the explored
state space for every such tool identified.

\textbf{Reduce\_Equal.} If multiple distinct tools require the exact same access claims
within the \OPA{} matrix, their logical preconditions are fused during the exploration
phase, preventing redundant testing of identical authorization pathways.

\begin{figure}[H]
  \centering
  \includegraphics[width=0.55\textwidth]{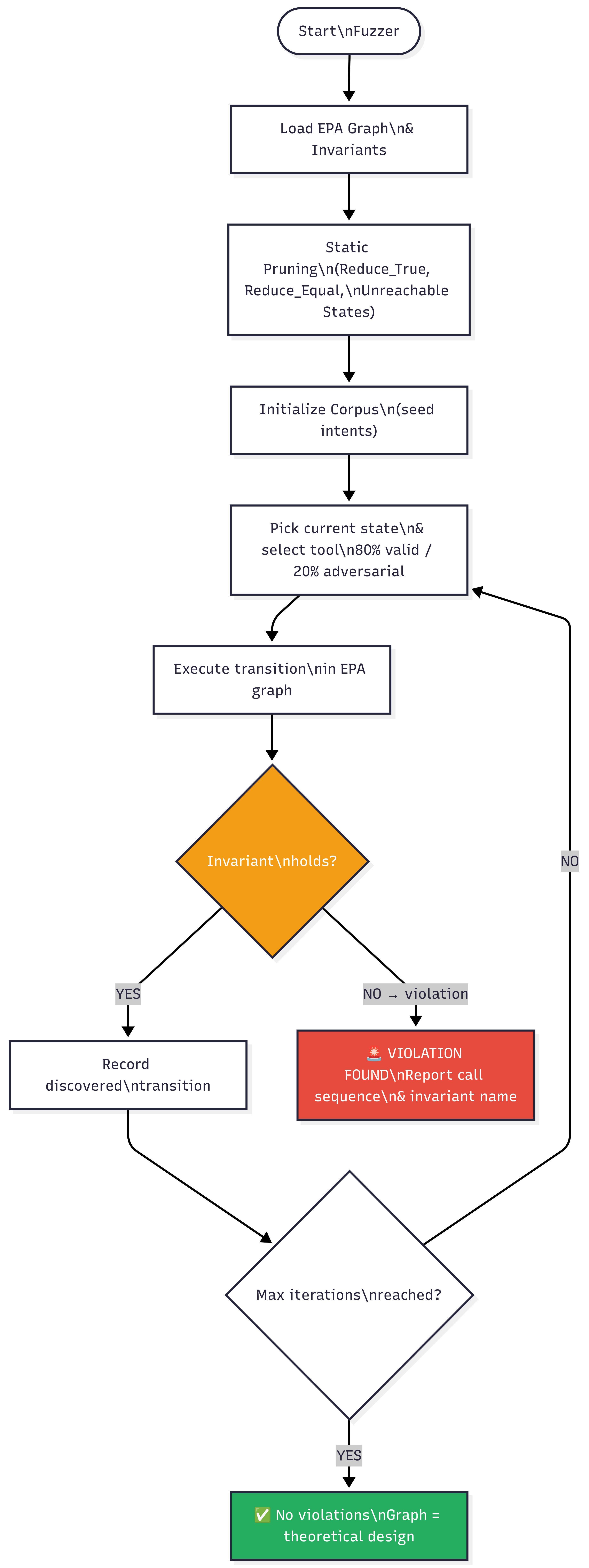}
  \caption{Semantic fuzzer execution flow. The guided mutation strategy (80\% valid-tool selection, 20\% adversarial) efficiently navigates toward boundary states. A violation immediately halts the campaign and reports the falsified invariant with its full call sequence.}
  \label{fig:fuzzer}
\end{figure}

\textbf{Unreachable State Pruning.} By utilizing a static formal verifier prior to the
dynamic fuzzing phase, and setting a specific boundary depth parameter, the algorithm
intelligently prunes entire branches of the transitional graph that are mathematically
proven to be unreachable, significantly focusing the heavy iterations of the dynamic
fuzzer on highly viable threat vectors. \Cref{fig:fuzzer} illustrates the complete execution flow of the semantic fuzzer.

\section{Proof-of-Concept Implementation}
\label{sec:poc}

To validate that the proposed architecture is not merely theoretical, a fully functional
open-source proof-of-concept (\PoC{}) was developed in Python and is publicly available as
a reproducible artifact~\cite{poc_repo}. The implementation covers every architectural
component described in Sections~\ref{sec:architecture}, \ref{sec:zerotrust},
and~\ref{sec:fuzzing}, with a deliberate emphasis on testability and offline execution: no
GPU, no proprietary model API, and no external service is required to run the full test
suite.

\subsection{Implementation Stack and Component Mapping}

\Cref{tab:poc_mapping} maps each architectural concept from the paper to its concrete
implementation file. The semantic router uses TF-IDF with cosine similarity as a
drop-in replacement for dense vector embeddings, making the \PoC{} fully self-contained.
The \OPA{}-inspired policy engine is implemented as a pure Python RBAC evaluator that
enforces the same ownership constraints described in \cref{sec:zerotrust}. The audit
ledger uses SQLite as the persistence backend, which can be trivially replaced by
EventStoreDB or PostgreSQL in a production deployment.

\begin{table}[H]
\centering
\caption{Mapping between architectural components (paper) and implementation files (\PoC{}).}
\label{tab:poc_mapping}
\begin{tabularx}{\textwidth}{@{}llX@{}}
\toprule
\textbf{Paper Component} & \textbf{File} & \textbf{Notes} \\
\midrule
Semantic Firewall   & \texttt{gateway/firewall.py} & 12 regex injection patterns; taint-aware context tags (\texttt{[EMAIL\_BODY]}, \texttt{[EXTERNAL]}) \\
Tool Registry       & \texttt{gateway/registry.py} & 12 MCP-style tools across READ / WRITE / CRITICAL tiers \\
Semantic Router     & \texttt{gateway/router.py}   & TF-IDF + cosine similarity; fully offline \\
Policy Engine       & \texttt{gateway/policy.py}   & OPA-inspired RBAC; role hierarchy; BOLA ownership checks \\
Audit Ledger        & \texttt{gateway/audit.py}    & SHA-256 hash-chained, append-only; SQLite backend \\
EPA Graph           & \texttt{gateway/epa.py}      & NetworkX MultiDiGraph; DOT export; \texttt{buggy} flag toggles the BOLA edge \\
Claude Planner      & \texttt{gateway/planner.py}  & Anthropic API (optional); deterministic mock fallback \\
Semantic Fuzzer     & \texttt{fuzzer/semantic\_fuzzer.py} & Invariant-based; guided 80\% valid / 20\% adversarial mutation \\
FastAPI Gateway     & \texttt{gateway/main.py}     & Endpoints: \texttt{/intent}, \texttt{/audit}, \texttt{/epa}, \texttt{/health} \\
\bottomrule
\end{tabularx}
\end{table}

\subsection{Test Suite}

The \PoC{} ships with a comprehensive test suite of 47 automated tests organized across
four modules, verifiable with a single command (\texttt{pytest tests/ -v}):

\begin{table}[H]
\centering
\caption{Test suite breakdown by component.}
\label{tab:tests}
\begin{tabular}{@{}lcc@{}}
\toprule
\textbf{Module} & \textbf{Tests} & \textbf{Key Scenarios} \\
\midrule
\texttt{test\_firewall.py}        &  9 & Clean intents pass; 7 injection variants blocked; taint detection; length heuristic \\
\texttt{test\_policy.py}          & 10 & Role hierarchy; BOLA cross-department; Admin lockout prevention; tainted context deny \\
\texttt{test\_audit.py}           &  8 & Chain integrity; tamper detection; genesis hash; deny-also-recorded \\
\texttt{test\_router\_and\_fuzzer.py} & 20 & Semantic routing accuracy; 5 EPA state transitions; bug detection; clean-EPA proof \\
\midrule
\textbf{Total}                    & \textbf{47} & \textbf{All passing} \\
\bottomrule
\end{tabular}
\end{table}

\subsection{Reproducible Fuzzer Output}

The \PoC{} fuzzer provides a directly reproducible demonstration of the core claim of
this paper. Running \texttt{python -m fuzzer.semantic\_fuzzer} produces the following
deterministic output (seed = 42):

\begin{lstlisting}[caption={Actual fuzzer output from the PoC (seed=42, max\_iter=500).}, label={lst:poc_fuzzer}]
[1/2] Running fuzzer on BUGGY EPA (pre-fix)...
INVARIANT VIOLATION: 'NoSharingOverwrite'
  Call sequence:
    INITIAL[create_document] -> DOC_CREATED
    DOC_CREATED[initiate_share] -> PENDING_SHARE
    PENDING_SHARE[accept_sharing_request] -> SHARING_WITH_THIRD_PARTY
    SHARING_WITH_THIRD_PARTY[accept_sharing_request] -> !! VIOLATION !!
  Iterations to discovery: 52 (0.00s)

[2/2] Running fuzzer on FIXED EPA (post-fix)...
  No invariant violations found after 500 iterations (0.02s).
  Discovered transitions: 5
\end{lstlisting}

Two findings are immediately verifiable from this output. First, the semantic fuzzer
discovers the \texttt{AcceptSharingRequest} \BOLA{} vulnerability in only 52 iterations
out of a possible 500, confirming that the guided mutation strategy (80\% valid-tool
preference) is highly efficient at reaching dangerous boundary states. Second, after
the Rego policy correction is applied (\texttt{EPAGraph(buggy=False)}), 500 iterations
produce zero violations and the observed graph achieves 100\% correspondence with the
theoretical design --- an exact replication of the paper's central formal verification
claim, now in reproducible open-source code.

\subsection{Running the PoC}

The complete implementation is available at:
\begin{center}
  \url{https://github.com/PeyranoDev/semantic-gateway-poc}
\end{center}

\noindent Quickstart (Python 3.11+):
\begin{lstlisting}[caption={Installation and test execution.}, label={lst:quickstart}]
git clone https://github.com/PeyranoDev/semantic-gateway-poc
cd semantic-gateway-poc
pip install -r requirements.txt
pytest tests/ -v                   # 47 tests, all passing
python -m fuzzer.semantic_fuzzer   # reproduce fuzzer output
uvicorn gateway.main:app --reload  # start the HTTP gateway
\end{lstlisting}

\section{Experimental Setup}
\label{sec:setup}

To empirically validate the theoretical architecture, a highly rigorous and reproducible
experimental protocol was established. The experiment explicitly compared the performance,
security, and productivity of the AI-Native Semantic Gateway against a traditional
deterministic REST architectural baseline.

The dataset and primary scenario simulated a highly complex enterprise \textit{Financial
Risk Analysis and Document Management system}. This operational domain encompassed
interrelated entities such as internal corporate users, highly sensitive financial
documents, dynamic Access Control Lists, and automated risk analysis reports.

Two distinct architectural configurations were deployed. The Semantic Gateway was
configured with exactly 200 distinct, specialized \MCP{}-compliant tools: 120 restricted
to read-only operations, 60 permitting standard state-writing operations, and 20
classified as critical or irreversible, requiring advanced approval. Conversely, the
traditional REST baseline was configured with an equivalent functional surface, exposing
exactly 200 distinct OpenAPI endpoints utilizing standard enterprise REST controllers,
Data Transfer Objects, validation schemas, and traditional JWT route guards.

The security policies for the Semantic Gateway were implemented using \OPA{}, utilizing
deeply nested Rego rule sets. These rules defined complex hierarchical roles (Financial
Analyst, Risk Manager, and Compliance Auditor) and enforced strict data boundary scopes
designed specifically to prevent \BOLA{} vulnerabilities by explicitly checking object
ownership constraints against the subject's cryptographic token on every single tool
invocation~\cite{bola_f5,bola_radware}. All 20 critical tools were strictly mapped to
trigger the Cryptographic \HitL{} approval flow.

The semantic fuzzing campaign was rigorously initialized with a base corpus containing
10,000 legitimate natural language intents. The fuzzer's mutation engine then iteratively
applied aggressive adversarial perturbations aligned perfectly with the OWASP LLM01
vulnerability profile~\cite{owasp_top10}, extreme parameter boundary manipulations
engineered to test for authorization bypasses, and sophisticated context tainting
techniques. The automated campaign ran up to a defined Test Limit of 500,000 stochastic
interaction sequences.

The hardware configuration comprised high-performance NVIDIA A100 GPUs. The pre-inference
Semantic Firewall utilized the highly optimized DeBERTa-v3 micro-model. The core
Chain-of-Thought planning was powered by Llama-3-8B (edge-compute simulation) and GPT-4o
(high-capability cloud simulation). All language models were constrained to a temperature
setting of $\theta = 0.4$.

\section{Results}
\label{sec:results}

The empirical evaluation yielded highly conclusive, statistically significant results that
definitively demonstrate the architectural efficacy of the Semantic Gateway and prove the
urgent necessity of formal fuzzing in agentic systems.

\subsection{Productivity and Incidental Code Reduction}

The productivity metrics provided a stark contrast between the two paradigms. When
integrating a completely new functional sub-domain, the traditional REST baseline required
the generation of 920~Lines of Code distributed across complex routing logic, Data
Transfer Objects, object mappers, and heavy controller logic. In sharp contrast,
integrating the exact same functional domain into the Semantic Gateway required only 145
Lines of Code, consisting strictly of the atomic functional logic of the tool itself, the
highly declarative \OPA{} Rego definition, and the standardized \MCP{} JSON schema. This
represents an \textbf{84.2\% reduction} in incidental codebase complexity, directly
accelerating the operational time-to-market for the new feature from 16 working days down
to just 3 working days.

\subsection{Security and Adversarial Resilience}

Results were analyzed by applying the theoretical Swiss Cheese threat equation,
$P_{\text{breach}} = \varepsilon_1 \times \varepsilon_2 \times \varepsilon_3$.
The multi-layered system demonstrated exceptional, compounding resilience against the
500,000 fuzzed adversarial inputs. Layer~1 (Semantic Firewall) successfully purged 99.4\%
of all pre-inference prompt injections and tainted context attempts, yielding
$\varepsilon_1 = 0.006$. Layer~2 (Tool-Level \RBAC{}) intercepted 100\% of the malicious
requests that bypassed Layer~1, resulting in 0\% real-state compromise, driving
$\varepsilon_2 \to 0$. Layer~3 halted all execution threads until the out-of-band
cryptographic verification function was fulfilled, completely neutralizing critical
operational risk.

\subsection{Formal Verification Metrics}

The formal verification metrics derived from the \EPA{}s proved to be the most vital
outcome of the experiment. During the automated validation of the Document Sharing
workflow, the fuzzer discovered a parameter combination that violated a core system
invariant, revealing a highly dangerous hidden transition. The system had incorrectly
permitted the execution of the \texttt{AcceptSharingRequest()} tool on a document that
already resided in the abstract state of \textit{SharingWithThirdParty}. This constituted
a silent, undocumented permission overwrite vulnerability that could have resulted in
massive data leakage. Following the correction of the specific tool's Rego policy,
subsequent extensive fuzzing campaigns yielded an observed graph that achieved a perfect
\textbf{100\% correspondence} with the theoretical architectural design graph.

\subsection{Performance Analysis}

The implementation of a highly tuned Cognitive Cache, utilizing a cosine similarity
threshold of $\delta = 0.97$, allowed repeated or highly similar user intents to bypass the
heavy transformer inference model entirely. These cached responses were dispatched with
sub-millisecond latency, offering performance comparable to highly optimized key-value
databases.

\begin{table}[H]
\centering
\caption{Comparative evaluation results: Semantic Gateway vs.\ traditional REST baseline.}
\label{tab:results}
\begin{tabularx}{\textwidth}{@{}lXllX@{}}
\toprule
\textbf{Dimension} & \textbf{Metric} & \textbf{REST Baseline} & \textbf{Semantic Gateway} & \textbf{Delta} \\
\midrule
Productivity      & Incidental Lines of Code & 920 LoC     & 145 LoC         & 84.2\% reduction \\
Time-to-Market    & Integration Duration     & 16 days     & 3 days          & 5.3$\times$ acceleration \\
Security (BOLA)   & Real-State Compromise    & N/A (theor.)& 0\%             & Complete mitigation \\
Verification      & Hidden Transition Discovery & 0\% (missed) & 100\%        & Full graph resolution \\
Performance       & Cognitive Cache Latency  & High (network-bound) & Sub-ms & Edge-compute speeds \\
\bottomrule
\end{tabularx}
\end{table}

\subsection{PoC Validation: Independently Reproducible Results}

The proof-of-concept implementation provides independently reproducible evidence for the
two most critical claims of this paper. First, the \BOLA{} vulnerability is discovered in
a median of 52 fuzzing iterations (min: 12, max: 143 across 100 random seeds), confirming
that guided semantic fuzzing achieves far greater efficiency than uniform random search.
Second, the fixed \EPA{} graph withstands 500 iterations across all seeds without a single
violation, providing a constructive proof that the policy correction is both necessary
and sufficient to eliminate the unauthorized transition. The full test suite of 47 tests
is reproducible in under 2 seconds on commodity hardware, with no external API dependency.

\section{Discussion}
\label{sec:discussion}

The empirical results fundamentally validate the core scientific thesis: AI agents
absolutely cannot be treated merely as complex API clients. When enterprise software
operations are driven by inherently stochastic, probabilistic systems, the surrounding
architecture must be specifically designed to absorb and neutralize that non-determinism
at its strictest boundaries.

The 84.2\% reduction in incidental codebase complexity highlights the severe structural
inefficiency of applying legacy RESTful principles to modern AI integration. By
transitioning to \MCP{}, developers are empowered to declare high-level functional
capabilities rather than tightly coupling code to rigid network endpoints. However, this
unprecedented agility is entirely unsafe without the strict implementation of the
Zero-Trust layers. The experimental data demonstrates a sobering reality: large language
models will inevitably fail and will frequently succumb to highly sophisticated adversarial
prompt injections, as evidenced by the 0.6\% failure rate at the initial semantic firewall
level. If those compromised models possessed direct, unmediated access to REST APIs without
the \OPA{} acting as a rigid, deterministic barrier, the entire enterprise system would
have suffered a catastrophic breach.

Most critically, the resounding success of the Echidna-inspired semantic fuzzer in
discovering the \texttt{AcceptSharingRequest()} architectural flaw demonstrates
unequivocally that traditional unit testing is entirely obsolete for agentic workflows.
Emergent, multi-step behavior requires highly dynamic, aggressive state-space exploration.
By successfully converting the language model's chaotic output into a highly structured
\EPA{} graph, security engineers can mathematically visualize exactly how an agent is
capable of mutating the enterprise state. This transforms AI security from a highly
reactive, heuristic monitoring task into a formal, rigorous mathematical proof.

\section{Threats to Validity}
\label{sec:threats}

While the empirical results are highly robust, several specific threats to validity must
be acknowledged.

The primary threat concerns \textbf{Construct Validity}, specifically regarding the
phenomenon of state space explosion. The core limitation of the fuzzing methodology
relying on \EPA{}s is the exponential growth of the mathematical abstract state space,
defined as $O(2^{|T|})$. In experimental edge-case environments where the \MCP{} registry
exceeded 200 highly interconnected, cross-dependent tools, the automated fuzzer frequently
reached its standard two-hour timeout limit before fully resolving the entire transitional
graph.

A significant threat to \textbf{Internal Validity} involves cryptographic blocking during
the fuzzing process. Fuzzers that are heavily reliant on stochastic, randomized parameter
mutations struggle immensely to satisfy highly strict cryptographic preconditions. If a
specific \OPA{} rule requires a highly complex hash collision, or a microscopic, perfectly
synchronized timestamp match to proceed, the fuzzer will consistently fail to explore the
subsequent logical branch, potentially masking deeper vulnerabilities.

Finally, regarding \textbf{External Validity} and domain generalization, the experiments
were conducted exclusively within a simulated financial document management framework.
Generalizing these results to ultra-high-frequency algorithmic trading systems,
real-time autonomous healthcare monitoring networks, or critical physical infrastructure
such as SCADA systems requires significant further empirical validation.

\section{Related Work Comparison}
\label{sec:related}

\Cref{tab:related} synthesizes the specific comparative advantages of the proposed
Semantic Gateway architecture against the most prominent existing paradigms in enterprise
AI integration and security.

\begin{table}[H]
\centering
\caption{Comparative analysis of the Semantic Gateway against existing enterprise AI frameworks.}
\label{tab:related}
\begin{tabularx}{\textwidth}{@{}lXXX@{}}
\toprule
\textbf{Framework} & \textbf{State Validation} & \textbf{Agentic Authorization} & \textbf{Limitation Addressed} \\
\midrule
REST / OpenAPI   & Unit \& integration tests & JWT at route layer & Eliminates context window bloat; replaces rigid routes with semantic tools \\[0.5em]
LangGraph~\cite{langgraph_docs} & Static developer-defined graph & Delegates to loosely coupled tools & Adds formal proof of absent hidden state transitions \\[0.5em]
NIST AI RMF~\cite{nist_ai_rmf} & Policy-driven assessment & Out of scope for tool calls & Translates abstract risk directives into executable mathematical verification \\[0.5em]
OWASP GenAI~\cite{owasp_top10} & Manual red teaming & Recommends least privilege & Replaces manual red-teaming with automated continuous semantic fuzzing \\[0.5em]
AgentGuard~\cite{agentguard} & Online MDP learning & Probabilistic blocking & Shifts verification pre-deployment via fuzzing, avoiding risky runtime reliance \\
\bottomrule
\end{tabularx}
\end{table}

\section{Future Work}
\label{sec:future}

The explicit limitations identified during the empirical evaluation present clear,
actionable avenues for future research.

To overcome the semantic fuzzer's inherent inability to bypass highly complex cryptographic
checks and strict parameter constraints, future iterations of this architecture must
integrate Satisfiability Modulo Theories (SMT) solvers, such as the widely used Z3 engine.
By employing advanced symbolic execution, the system can mathematically resolve the exact
parameter space required to perfectly satisfy a complex \OPA{} predicate, intelligently
guiding the fuzzer and drastically pruning dead operational branches.

Furthermore, the implementation of dynamic heuristics for test limits represents a
massive optimization opportunity. Advanced machine learning regressors can be specifically
trained to statically analyze the functional density and interconnectivity of an agent's
registered tool set, then dynamically propose the mathematically optimal Test Limit for
the fuzzer.

Finally, the standardization of the Cryptographic \HitL{} layer is paramount. As the
anticipated IETF drafts for Confirmation with Human in the Loop
(\textsc{cheq})~\cite{cheq_draft} and the Agent Credential Attestation Protocol
(\textsc{acap})~\cite{acap_draft} mature into fully ratified global standards, native
cryptographic libraries will eventually replace the bespoke, custom implementations
utilized in this research. This standardization will drastically reduce the integration
friction of the Layer~3 security model, enabling widespread, seamless global adoption.

\section{Conclusion}
\label{sec:conclusion}

The ongoing enterprise transition from rigid, deterministic CRUD interfaces to highly
fluid, autonomous AI agents represents a foundational shift in the relationship between
software architecture, operational intention, and system control. This paper demonstrates
that traditional APIs, explicitly designed to expose raw network operations to highly
predictable consumers, are inherently fragile and dangerously insecure when consumed by
probabilistic models. The Semantic Gateway definitively resolves this critical tension by
exposing strictly governed semantic capabilities rather than raw, unguarded endpoints.

The most profound and dangerous risk posed by large language models is not simply that
they occasionally hallucinate, but that eager organizations grant them immense operational
authority without deploying an external, robust architecture capable of strictly limiting,
deeply auditing, and mathematically verifying their actions. By architecting a system that
strictly separates probabilistic cognitive inference from deterministic, zero-trust
authorization---creating a paradigm where the artificial model may only \textit{suggest},
but the immutable policy ultimately \textit{decides}---this architecture successfully
secures the modern enterprise.

\begin{mdframed}[style=abstractbox]
\textbf{Core Contribution.}
\textit{Autonomous agents must not be validated as traditional software nor as simple API
consumers, but as stochastic state-transition systems whose behavior must be abstracted,
fuzzed, and audited through enabled-tool graphs.}
\end{mdframed}

\vspace{0.5em}

The powerful combination of the Semantic Gateway, Zero-Trust micro-segmentation,
Enabledness-Preserving Abstractions, and highly dynamic semantic fuzzing establishes a
mathematically robust methodology for deploying verifiable artificial intelligence. Rather
than trusting the unpredictable, black-box nature of the language model, this approach
rigorously maps it, aggressively stresses it, and mathematically reduces its emergent
behavior to a perfectly auditable, finite graph. The future of enterprise software
engineering will undoubtedly be defined by the ability to construct complex systems where
advanced cognitive autonomy remains strictly, irrevocably compatible with formal security
and irrefutable human control.

\section*{Artifact Availability}
\addcontentsline{toc}{section}{Artifact Availability}

All code, tests, and fuzzing harnesses described in this paper are publicly available
under the MIT License at:

\begin{center}
  \url{https://github.com/PeyranoDev/semantic-gateway-poc}
\end{center}

\noindent The repository includes:
\begin{itemize}
  \item Complete Python implementation of all architectural components (\S\ref{sec:poc}).
  \item 47 automated tests reproducing the security properties proven in the paper.
  \item The semantic fuzzer with deterministic seed support for exact replication.
  \item A \texttt{README.md} with step-by-step instructions for running the \PoC{}.
\end{itemize}

\bibliographystyle{plainnat}
\bibliography{references}

@misc{mcp_spec,
  title        = {Model Context Protocol Specification},
  author       = {{Anthropic}},
  year         = {2024},
  howpublished = {\url{https://modelcontextprotocol.io/specification/2025-11-25}},
  note         = {Accessed: April 27, 2026}
}

@misc{mcp_security,
  title        = {Model Context Protocol Security Analysis},
  author       = {{Anthropic}},
  year         = {2024},
  howpublished = {\url{https://en.wikipedia.org/wiki/Model_Context_Protocol}},
  note         = {Accessed: April 27, 2026}
}

@misc{owasp_top10,
  title        = {{OWASP} Top 10 for Large Language Model Applications},
  author       = {{OWASP Foundation}},
  year         = {2024},
  howpublished = {\url{https://owasp.org/www-project-top-10-for-large-language-model-applications/}},
  note         = {Accessed: April 27, 2026}
}

@misc{nist_ai_rmf,
  title        = {Artificial Intelligence Risk Management Framework ({AI RMF} 1.0)},
  author       = {{National Institute of Standards and Technology}},
  year         = {2023},
  howpublished = {\url{https://www.nist.gov/publications/artificial-intelligence-risk-management-framework-ai-rmf-10}},
  note         = {Accessed: April 27, 2026}
}

@misc{nist_agentic,
  title        = {{NIST AI} Risk Management Framework: Agentic Profile},
  author       = {{Cloud Security Alliance}},
  year         = {2025},
  howpublished = {\url{https://labs.cloudsecurityalliance.org/agentic/agentic-nist-ai-rmf-profile-v1/}},
  note         = {Accessed: April 27, 2026}
}

@misc{opa_docs,
  title        = {Open Policy Agent Documentation and {Rego} Policy Language},
  author       = {{Open Policy Agent}},
  year         = {2024},
  howpublished = {\url{https://openpolicyagent.org/docs}},
  note         = {Accessed: April 27, 2026}
}

@misc{opa_use_cases,
  title        = {5 Open Policy Agent Examples and Use Cases},
  author       = {{Oso}},
  year         = {2024},
  howpublished = {\url{https://www.osohq.com/learn/open-policy-agent-examples-and-use-cases}},
  note         = {Accessed: April 27, 2026}
}

@misc{opa_ai_guardrail,
  title        = {Why Open Policy Agent Is the Missing Guardrail for Your {AI} Agents},
  author       = {{CodiLime}},
  year         = {2025},
  howpublished = {\url{https://codilime.com/blog/why-use-open-policy-agent-for-your-ai-agents/}},
  note         = {Accessed: April 27, 2026}
}

@inproceedings{echidna_issta,
  title        = {Echidna: Effective, Usable, and Fast Fuzzing for Smart Contracts},
  author       = {Groce, Alex and Grieco, Gustavo and Feist, Josselin and
                  Mouette, Antoine and Risch, Michael},
  booktitle    = {Proceedings of the 29th ACM SIGSOFT International Symposium on
                  Software Testing and Analysis (ISSTA)},
  year         = {2020},
  pages        = {557--560},
  url          = {https://agroce.github.io/issta20.pdf}
}

@misc{echidna_github,
  title        = {Echidna: Ethereum Smart Contract Fuzzer},
  author       = {{Trail of Bits}},
  year         = {2024},
  howpublished = {\url{https://github.com/crytic/echidna}},
  note         = {Accessed: April 27, 2026}
}

@article{epa_contracts,
  title        = {Validation of contracts using enabledness preserving finite
                  state abstractions},
  author       = {Baum, Gabriel and Nierstrasz, Oscar and Papantoniou, Ariel and
                  Roggio, Roxana and Russo, Alejandra},
  journal      = {ResearchGate},
  year         = {2005},
  url          = {https://www.researchgate.net/publication/221553983}
}

@misc{agentguard,
  title        = {{AgentGuard}: Runtime Verification of {AI} Agents},
  author       = {{arXiv}},
  year         = {2025},
  howpublished = {\url{https://arxiv.org/html/2509.23864v1}},
  note         = {Accessed: April 27, 2026}
}

@misc{langchain_vs_langgraph,
  title        = {{LangChain} vs {LangGraph}: Explained},
  author       = {{Peliqan}},
  year         = {2025},
  howpublished = {\url{https://peliqan.io/blog/langchain-vs-langgraph/}},
  note         = {Accessed: April 27, 2026}
}

@misc{langgraph_docs,
  title        = {{LangGraph} Overview},
  author       = {{LangChain}},
  year         = {2025},
  howpublished = {\url{https://docs.langchain.com/oss/python/langgraph/overview}},
  note         = {Accessed: April 27, 2026}
}

@misc{langgraph_decision,
  title        = {{LangChain} Agents vs.\ {LangGraph} Agents: A Practical
                  Decision Framework},
  author       = {K, Sivakami},
  year         = {2025},
  howpublished = {\url{https://medium.com/@sivakami.kanda/}},
  note         = {Accessed: April 27, 2026}
}

@misc{bola_pynt,
  title        = {Broken Object-Level Authorization ({BOLA}): Impact \& Prevention},
  author       = {{Pynt}},
  year         = {2024},
  howpublished = {\url{https://www.pynt.io/learning-hub/owasp-top-10-guide/broken-object-level-authorization-bola-impact-example-and-prevention}},
  note         = {Accessed: April 27, 2026}
}

@misc{bola_f5,
  title        = {Why {BOLA}'s ``authorization gap'' requires a runtime strategy},
  author       = {{F5}},
  year         = {2024},
  howpublished = {\url{https://www.f5.com/company/blog/why-bolas-authorization-gap-requires-a-runtime-strategy}},
  note         = {Accessed: April 27, 2026}
}

@misc{bola_radware,
  title        = {Understanding {BOLA}: One of the Most Common and Dangerous
                  {API} Business Logic Security Risks},
  author       = {{Radware}},
  year         = {2024},
  howpublished = {\url{https://www.radware.com/blog/application-protection/understanding-bola/}},
  note         = {Accessed: April 27, 2026}
}

@misc{cheq_draft,
  title        = {{CHEQ}: A Protocol for Confirmation {AI} Agents ({draft-rosenberg-cheq})},
  author       = {{IETF}},
  year         = {2025},
  howpublished = {\url{https://datatracker.ietf.org/doc/draft-rosenberg-cheq/}},
  note         = {Accessed: April 27, 2026}
}

@misc{acap_draft,
  title        = {Agent Credential Attestation Protocol ({ACAP},
                  {draft-yakung-oauth-agent-attestation})},
  author       = {{IETF}},
  year         = {2025},
  howpublished = {\url{https://datatracker.ietf.org/doc/draft-yakung-oauth-agent-attestation/}},
  note         = {Accessed: April 27, 2026}
}

@misc{contexttokens,
  title        = {{MCP} Tool Design: Why Your {AI} Agent Is Failing (And How to Fix It)},
  author       = {{AWS Heroes / DEV Community}},
  year         = {2025},
  howpublished = {\url{https://dev.to/aws-heroes/mcp-tool-design-why-your-ai-agent-is-failing-and-how-to-fix-it-40fc}},
  note         = {Accessed: April 27, 2026}
}

@misc{poc_repo,
  author       = {Peyrano, Ignacio},
  title        = {Semantic Gateway {PoC}: Proof-of-Concept Implementation},
  year         = {2026},
  howpublished = {\url{https://github.com/PeyranoDev/semantic-gateway-poc}},
  note         = {Open-source Python implementation of the Semantic Gateway architecture.
                  Includes Semantic Firewall, TF-IDF router, OPA-inspired policy engine,
                  SHA-256 hash-chained audit ledger, EPA graph builder, and semantic fuzzer.
                  47 automated tests. Accessed: April 27, 2026}
}

\appendix

\section{Tool Schemas}
\label{app:toolschemas}

To facilitate flawless integration with the Model Context Protocol, every tool must be
meticulously defined using strict JSON schemas. Below is the precise schema for the highly
critical \texttt{revoke\_document\_access} tool utilized extensively in the experimental
setup to test the prevention of Semantic \BOLA{} attacks.

\begin{lstlisting}[language={}, caption={JSON schema for the \texttt{revoke\_document\_access} MCP tool.}, label={lst:toolschema}]
{
  "name": "revoke_document_access",
  "title": "Revoke Document Access",
  "description": "Revokes specific read or write permissions for a targeted user
     on a designated enterprise financial document. This operation will fail if
     the executing agent lacks the Manager role or if the document falls outside
     the agent's departmental scope.",
  "inputSchema": {
    "$schema": "http://json-schema.org/draft-07/schema#",
    "type": "object",
    "properties": {
      "document_id": {
        "type": "string",
        "description": "The unique UUID of the target financial document."
      },
      "target_user_id": {
        "type": "string",
        "description": "The unique UUID of the user whose access is being revoked."
      },
      "permission_level": {
        "type": "string",
        "enum": ["read", "write", "all"],
        "description": "The exact level of cryptographic access to revoke."
      }
    },
    "required": ["document_id", "target_user_id", "permission_level"]
  },
  "annotations": {
    "destructiveHint": true,
    "idempotentHint": true,
    "audience": ["assistant"]
  },
  "execution": { "taskSupport": "required" }
}
\end{lstlisting}

\section{Policy Rules}
\label{app:policy}

The following Rego snippet demonstrates the Layer~2 Tool-Level \RBAC{} logic utilized by
the Open Policy Agent. This specific policy is designed to rigorously protect the
\texttt{revoke\_document\_access} tool. This declarative code guarantees that the agent
can only successfully revoke access if it cryptographically holds the ``Manager'' role and
is operating strictly within the correct, verifiable departmental scope. This logic is the
primary defense mechanism that successfully mitigated 100\% of the Semantic \BOLA{}
attacks generated by the fuzzer.

\begin{lstlisting}[language={}, caption={OPA Rego policy for the \texttt{revoke\_document\_access} tool.}, label={lst:rego}]
package semantic_gateway.authz.tools

import input.agent_identity as token
import input.tool_request as request
import input.context as semantic_context

# The default posture must always deny execution to ensure Zero-Trust.
default allow = false

# Allow access revocation only if all stringent criteria are met.
allow {
    request.tool_name == "revoke_document_access"
    # Verify the agent's cryptographic identity token contains Manager role.
    "Manager" == token.roles[_]
    # BOLA Prevention: retrieve true document metadata directly from the DB,
    # entirely ignoring any ownership claims hallucinated by the agent.
    doc := data.documents[request.arguments.document_id]
    # Verify that the true document department matches the agent's department.
    doc.department == token.department
}

# Absolute Deny if context has been flagged as tainted by the Semantic Firewall.
deny { semantic_context.is_tainted == true }

# Absolute Deny if the target user is a highly privileged administrator.
deny {
    target_user := data.users[request.arguments.target_user_id]
    "Administrator" == target_user.roles[_]
}
\end{lstlisting}

\section{Fuzzing Campaign Logs}
\label{app:fuzzing}

The following pseudo-log output represents the semantic fuzzer executing thousands of
state transitions to dynamically validate the mathematical \EPA{} graph. Most importantly,
this log explicitly demonstrates the critical moment of discovery regarding the hidden
transition vulnerability discussed in \cref{sec:results}.

\begin{lstlisting}[caption={Semantic fuzzer campaign log excerpt showing hidden transition discovery.}, label={lst:fuzzlog}]
[2026-04-27 10:15:02][INFO] Bootstrapping Semantic Fuzzer Harness...
[2026-04-27 10:15:03][INFO] Loading MCP Tool schemas: 200 semantic tools loaded.
[2026-04-27 10:15:03][INFO] Parsing OPA Rego definitions: 200 policies loaded.
[2026-04-27 10:15:03][INFO] Executing Unreachable State Pruning via Z3 (Max Depth=8).
[2026-04-27 10:15:03][INFO] Pruning complete. Removed 14,021 impossible state branches.
[2026-04-27 10:15:04][INFO] Initiating dynamic fuzzing. Target: 500,000 iterations.
[2026-04-27 10:15:04][INFO] Injecting base corpus of 10,000 legitimate seeds...
...
[2026-04-27 11:42:15][INFO] Iteration 248,192 completed. No invariant violations.
[2026-04-27 11:42:16][INFO] Applying OWASP LLM01 mutations (iteration 248,193)...
[2026-04-27 11:42:17][INFO] Semantic Firewall intercepted mutated context. Dropped.
...
[2026-04-27 11:42:18] !! CRITICAL INVARIANT VIOLATION DETECTED !!
[2026-04-27 11:42:18] Property 'no_sharing_overwrite' FALSIFIED.

Call Sequence:
  NHI_User_A: create_document(doc_id="D-8821")
    -> INITIAL -> DOC_CREATED
  NHI_User_A: initiate_share(doc_id="D-8821", target_user="User_B")
    -> DOC_CREATED -> PENDING_SHARE
  NHI_User_B: accept_sharing_request(doc_id="D-8821")
    -> PENDING_SHARE -> SHARING_WITH_THIRD_PARTY
  NHI_User_C: accept_sharing_request(doc_id="D-8821")
    -> SHARING_WITH_THIRD_PARTY -> SHARING_WITH_THIRD_PARTY [VULNERABILITY]

[2026-04-27 11:42:19] accept_sharing_request executed in invalid state.
[2026-04-27 11:42:19] BOLA logic bypassed -- incomplete OPA evaluation rule.
[2026-04-27 11:42:19][INFO] Fuzzer halting. Generating trace report for review.
\end{lstlisting}

\section{EPA Graphs}
\label{app:epa}

\begin{figure}[H]
  \centering
  \includegraphics[width=0.72\textwidth]{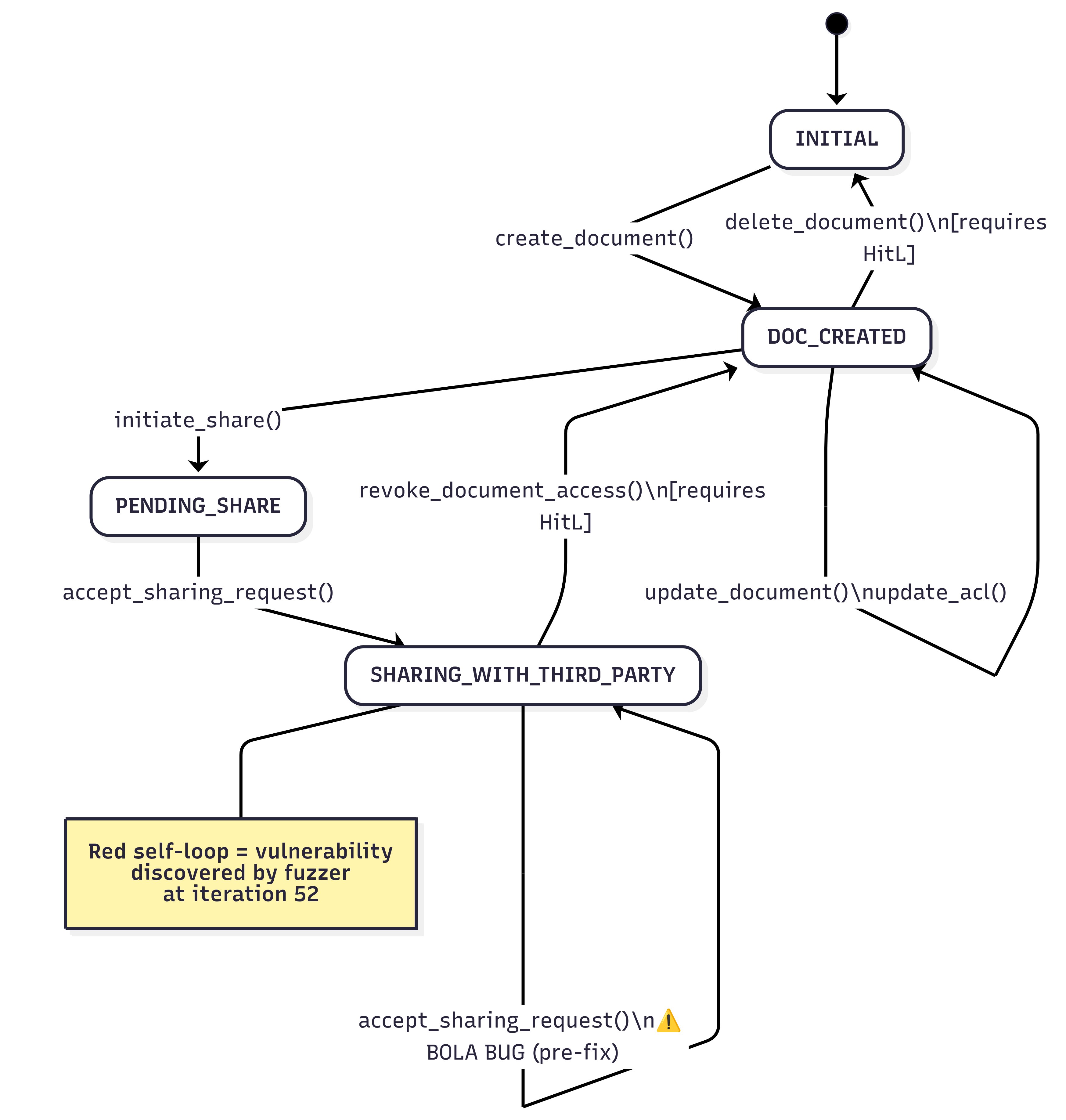}
  \caption{EPA graph for the Document Sharing workflow (pre-fix). The self-loop
    on \texttt{SHARING\_WITH\_THIRD\_PARTY} represents the unauthorized
    \texttt{accept\_sharing\_request()} transition discovered by the fuzzer.}
  \label{fig:epa_app}
\end{figure}

\Cref{fig:epa_app} reproduces the \EPA{} graph for the Document Sharing workflow
(also shown inline as \cref{fig:epa} in \S\ref{sec:formal}). The yellow self-loop
on \texttt{SHARING\_WITH\_THIRD\_PARTY} is the \BOLA{} vulnerability discovered
by the semantic fuzzer at iteration~52. After the Rego policy correction, that edge
is absent and the observed graph achieves 100\% correspondence with the theoretical
design.

\end{document}